# Dislocation interactions in olivine control postseismic creep of the upper mantle

David Wallis[1†*], Lars N. Hansen[2], Angus J. Wilkinson[3], Ricardo A. Lebensohn[4]

[1]Department of Earth Sciences, Utrecht University, 3584 CB Utrecht, The Netherlands.
[2]Department of Earth Science, University of Minnesota-Twin Cities, Minneapolis, Minnesota, U.S.A. 55455.
[3]Department of Materials, University of Oxford, Oxford, OX1 3PH, U.K.
[4]Los Alamos National Laboratory, Los Alamos, New Mexico, U.S.A.
[†]Present address: Department of Earth Sciences, University of Cambridge, Cambridge, CB2 3EQ, U.K.
*Corresponding author

## Abstract

Changes in stress applied to mantle rocks, such as those imposed by earthquakes, induce a period of evolution in viscosity and microstructure. This transient creep is often modelled based on stress transfer among slip systems due to grain interactions. However, recent experiments have demonstrated that the intragranular accumulation of stresses among dislocations is the dominant cause of strain hardening in olivine at low temperatures, raising the question of whether the same process contributes to transient creep at higher temperatures. Here, we demonstrate that olivine samples deformed at 25°C or 1150–1250°C both contain stress heterogeneities of ~1 GPa that are imparted by dislocations and have correlation lengths of ~1 μm. The similar stress distributions formed in both temperature regimes indicate that accumulation of stresses among dislocations also provides a contribution to transient creep at high temperatures. The results motivate a new generation of models that capture these intragranular processes and may refine predictions of evolving mantle viscosity over the earthquake cycle.

## Introduction

Major earthquakes impose changes in stress on the hotter, viscous rocks underlying the fault zone[1–7]. Relaxation of these stresses contributes to postseismic deformation and stress redistribution within and between fault zones over the earthquake cycle[1–7]. Modelling these processes is challenging because changes in the stress applied to viscous rocks induce a period of transient evolution in viscosity, which is observable in laboratory experiments[8–14] and detectable in geodetic datasets[1–4,15]. In both these contexts, the observations are often best fit by rheological models with nonlinear stress dependencies indicating that dislocation motion plays a dominant role[1–4,8–10]. Experiments on geological minerals[9,16–18] and metals[19,20] have demonstrated that this transient creep results from subtle changes in the dislocation microstructure and/or micromechanical state in response to a change in applied stress. Constraints on these underlying microphysical processes are therefore key to the formulation of rheological models that can be confidently extrapolated between laboratory and natural deformation conditions.

Previous studies have considered rheological behaviours typical of both low-temperature and high-temperature conditions to contribute to postseismic transient creep. Geological studies have focused on deformation microstructures in exhumed shear zones that are inferred to have directly underlain frictional faults, in which the stress changes and their associated microstructural expressions are most pronounced[6,21]. In these settings, postseismic deformation is recorded by dislocation microstructures (e.g., slip bands and cells of tangled dislocations) associated with low-temperature plasticity that is inferred to have been induced by transient increases in stress[6,21]. In contrast, postseismic geodetic signals can capture far-field deformation distributed throughout the lower crust and upper mantle, for which temperatures are typically higher and stress changes are smaller than directly adjacent to the seismogenic zone[1–4,15,22]. In these contexts, deformation by dislocation creep or dislocation-accommodated grain boundary sliding is likely commonplace[23–25].

Most quantitative descriptions of transient creep used to analyse postseismic geodetic data are largely phenomenological in that they describe observed material behaviour but lack a comprehensive basis in the underlying physical processes. A widely used example is the Burgers rheology[1,2,26–28], which is capable of generating a time-dependent evolution in viscosity, anelastic behaviour, and steady-state flow. Whilst attempts have been made to link a Burgers rheology to transient dislocation creep[1,2,28], the overarching arrangement of the Maxwell and Kelvin elements does not arise naturally from the fundamentals of dislocation motion.

A model with a deeper basis in the microphysics of dislocation creep has been proposed by Karato[29], inspired by earlier experiments on water ice[12,13]. In this model, transient creep arises from stress transfer between populations of grains with different crystallographic orientations. Upon loading, initial dislocation motion occurs in grains with high resolved shear stress on the weakest slip system. However, maintenance of strain compatibility in an aggregate of anisotropic grains requires activation of additional, stronger slip systems in other grains to counteract stress concentrations generated during progressive deformation. This progressive transition in the slip systems that control the bulk strain rate results in a progressive increase in viscosity[29].

Whilst the transfer of stress among slip systems in grains of different orientations does potentially contribute to transient creep of rocks, a variety of studies have demonstrated that transient creep also occurs even in single crystals, for which the stress-transfer model cannot apply. Strain hardening transients are exhibited by single crystals of olivine deforming by both low-temperature plasticity[30–32] and power-law creep at high temperatures[9,33,34], with the transition between mechanisms occurring at temperatures of approximately 1000–1100°C at typical experimental stresses and strain rates. Microstructural analyses of single crystals deformed in both temperature regimes indicate that strain hardening results from increases in dislocation density[9,35,36] and long-range interactions among dislocations via their stress fields[36,37]. These observations imply that these intragranular processes make an essential contribution to strain-hardening transients that has been largely overlooked. This suggestion is supported by a comparison between strain hardening behaviour of single crystals and that of aggregates of olivine deformed at room temperature, which are indistinguishable[31]. Therefore, it seems clear that transient creep of aggregates deformed in low-temperature plasticity is controlled by intragranular dislocation interactions. However, a major outstanding question remains. Do similar interactions among dislocations contribute to transient deformation at the high temperatures relevant to the regions of the lithosphere contributing to postseismic creep?

We hypothesise that the same intragranular processes control transient creep of aggregates of olivine deformed by power-law creep and those deformed by low-temperature plasticity. We suggest that samples deformed in either regime will share similarities in characteristics of their residual stress fields. Specifically, we test for the presence of intragranular stress heterogeneity, whether that heterogeneity can be attributed to the dislocation content, and whether that heterogeneity occurs over length-scales greater than the average dislocation spacing, indicating the stresses arise from long-range dislocation interaction.

We carry out these tests by analysing intragranular stress heterogeneity within olivine deformed at either 25°C[31] or 1150–1250°C[23] using high-angular resolution electron backscatter diffraction (HR-EBSD) (Methods). Unlike conventional EBSD, which struggles to resolve the subtle microstructural changes associated with transient creep at small strains[38], HR-EBSD provides exceptionally precise estimates of the density of geometrically necessary dislocations and, importantly, maps heterogeneity in elastic strain and residual stress[39–42]. We analyse the stress distributions in terms of the theory, established in the materials sciences[43–45], for stress fields of a population of dislocations to test the causality between stress heterogeneity and the dislocation content (Methods). In particular, we test whether tails of the probability (*P*) distributions of shear stress ($\sigma_{12}$) follow a $P(\sigma_{12}) \propto \sigma_{12}^{-3}$ relationship, as expected of stress fields generated by dislocations[43–46] (Methods). Autocorrelation of the stress fields provides a measure of the characteristic length scale of stress variation (Methods) and therefore a test for the presence of long-range internal stress. If the hypothesis is supported by our results, these observations will provide the basis for a new generation of rheological models of transient creep, rooted in the microphysics of intracrystalline deformation.

# Results

### *Geometrically necessary dislocation density*

Figure 1 presents maps of the densities of geometrically necessary dislocations in each sample. The undeformed single crystal, MN1, exhibits an apparent density of approximately $1\times10^{12}$ m$^{-2}$, resulting from noise in the rotation measurements[47]. This value is consistent with a total dislocation density of $< 10^{10}$ m$^{-2}$ measured previously by oxidation decoration[47]. The undeformed aggregate, PI-1523s, contains GND densities generally $< 10^{14}$ m$^{-2}$, but with some grains dissected by bands of GND densities $> 10^{14}$ m$^{-2}$ that mark subgrain boundaries. The single crystal deformed at room temperature, San382t, displays parallel linear arrays of elevated GND density, on the order of $10^{14}$ m$^{-2}$. In contrast, the aggregates deformed at room temperature, San382b and San372b, lack this linear structure and contain few regions with GND densities $< 10^{14}$ m$^{-2}$. The samples deformed at high temperatures, PI-1488, PI-1523, and PI-1519, contain GND densities broadly comparable to those of San382b but typically exhibit smoother variation in GND density within grains.

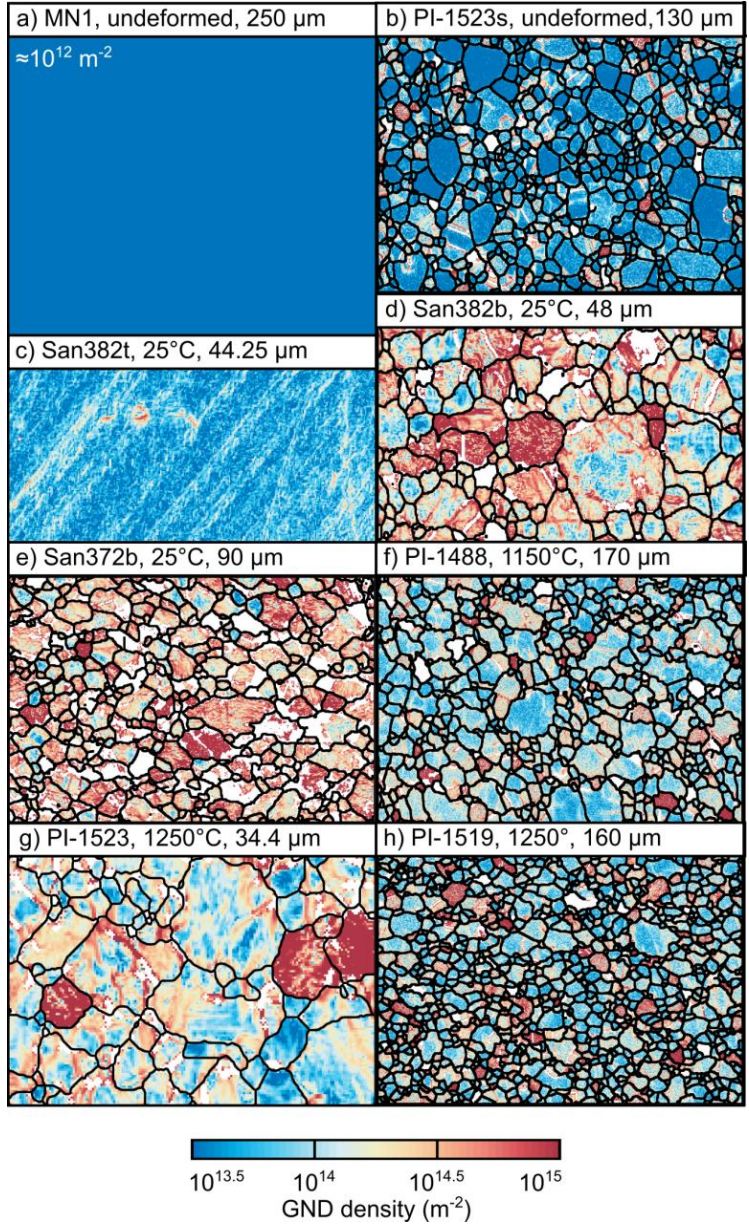

*Figure 1.* *Densities of geometrically necessary dislocations (GNDs) estimated from lattice rotations measured by HR-EBSD. (a) and (c) are single crystals, whereas all other maps are aggregates. Grains with apparent GND densities uniformly > $10^{15}$ $m^{-2}$ have orientations that result in high noise floors, obscuring the intragranular GND structures[39]. Annotations indicate the sample number, final deformation temperature, and length of the horizontal dimension. The compression direction was vertical in (c–h). White areas were not indexed during EBSD acquisition or failed the cross-correlation quality criteria. Black lines mark grain boundaries with misorientation angles > 10°.*

## *Spatial heterogeneity in residual stress*

Figure 2 presents maps of $\sigma_{12}$ normalised relative to the mean value within each grain (Methods). The single-crystal control sample, MN1, exhibits a homogeneous stress distribution with a standard deviation of 70 MPa. The isostatically hot-pressed aggregate, PI-1523s, exhibits stress heterogeneity that typically varies smoothly by a few hundred megapascals within grains, along with occasional stress concentrations. All other samples exhibit more pronounced intragranular stress heterogeneity. Sample San382t contains bands of elevated stress of alternating sign that vary in magnitude on the order of 1 GPa over distances of a few micrometres. The deformed aggregates exhibit stress distributions that are qualitatively similar to each other but lack the ordered structure displayed by San382t. In each case, stress typically varies smoothly between domains of stress on the order of 1 μm across, with alternating sign and magnitudes again on the order of 1 GPa.

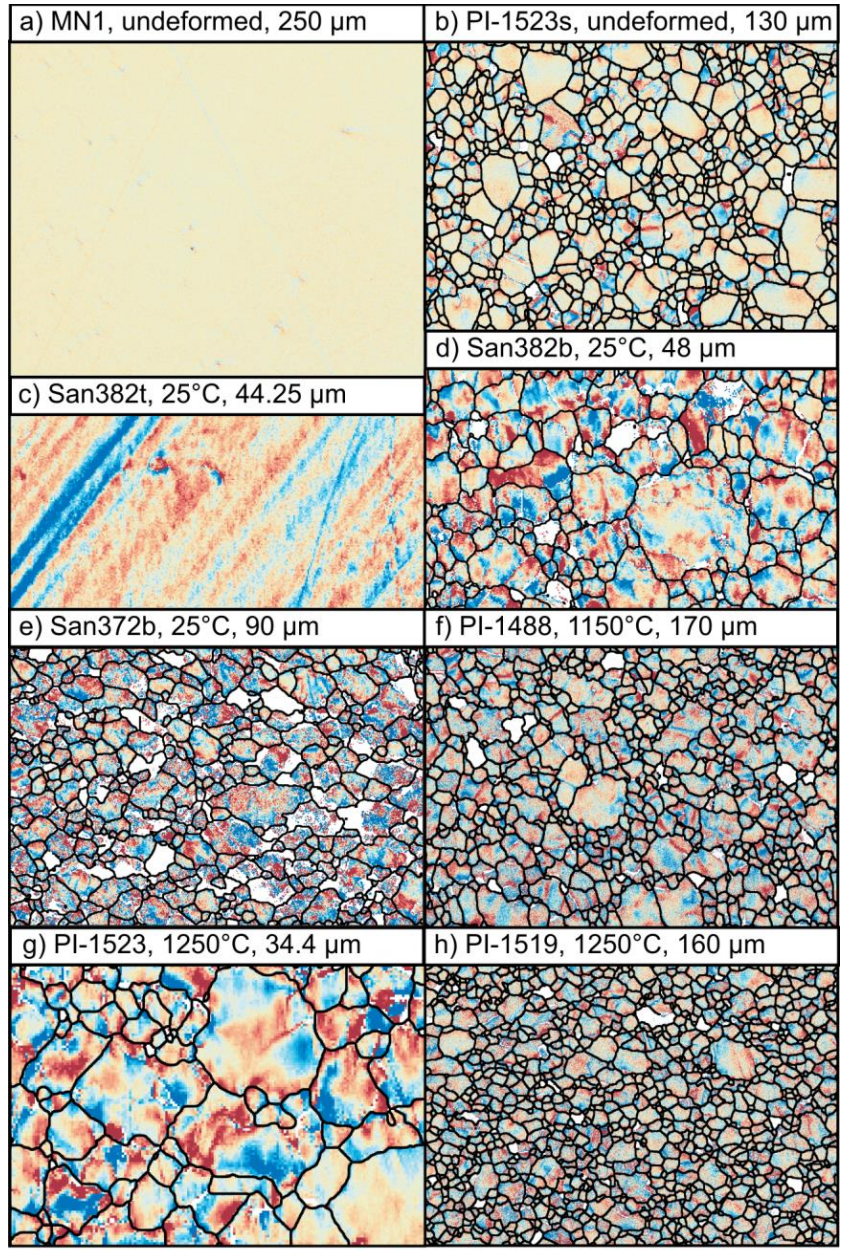

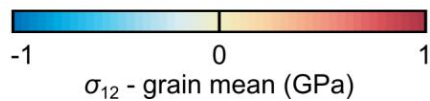

$\sigma_{12}$ - grain mean (GPa)

*Figure 2.* *Maps of $\sigma_{12}$ normalised to the mean value within each grain. Annotations indicate the sample number, final deformation temperature, and length of the horizontal dimension. The compression direction was vertical in (c–h). White areas were not indexed during EBSD acquisition or failed the cross-correlation quality criteria. Black lines mark grain boundaries with misorientation angles > 10°.*

## *Probability distributions of residual-stress heterogeneity*

Figure 3 presents probability distributions of normalised $\sigma_{12}$ in each sample. All the deformed samples exhibit markedly broader distributions than the corresponding undeformed control samples. The stress distributions are broader in the deformed aggregates than in the deformed single crystal but all deformed samples contain distributions that extend beyond ±1 GPa. The probability distributions of the aggregates deformed at low temperatures are similar to, or broader than, those of aggregates deformed at high temperatures.

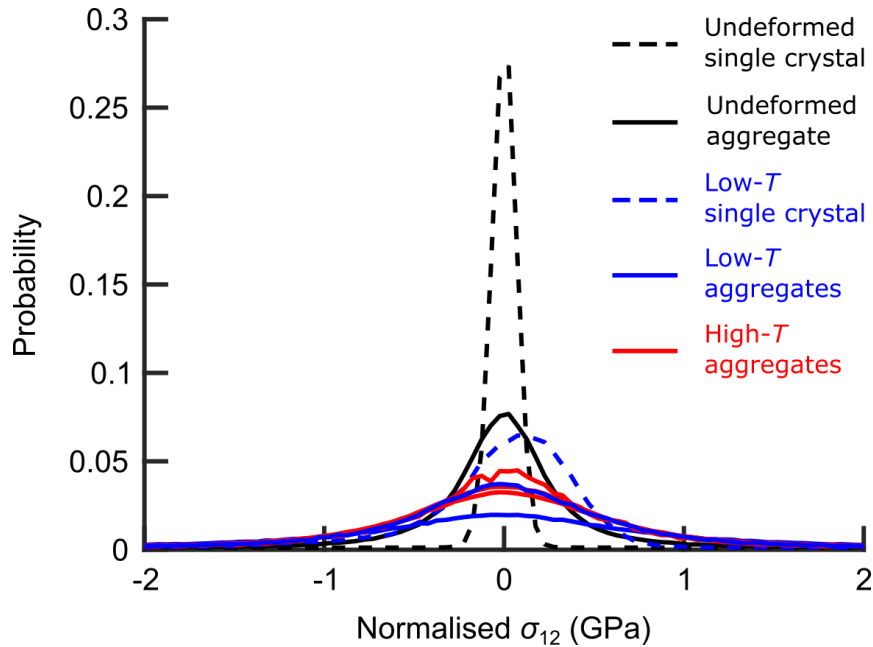

*Figure 3.* *Probability distributions of normalised $\sigma_{12}$ calculated using bin widths of 50 MPa.*

Figure 4 characterises the probability distributions of normalised $\sigma_{12}$ in each sample. The normal probability plot in Figure 4a demonstrates that the central, low-stress portion of each distribution falls on a straight line corresponding to a normal distribution (Methods). However, the figure reveals that the distribution for each sample exhibits a 'tail' that deviates from the straight line to higher stress magnitudes. In the deformed single crystal and aggregates, the magnitude of this deviation is greater than in the corresponding control samples. These tails correspond to stress heterogeneities typically greater than 1 GPa in magnitude.

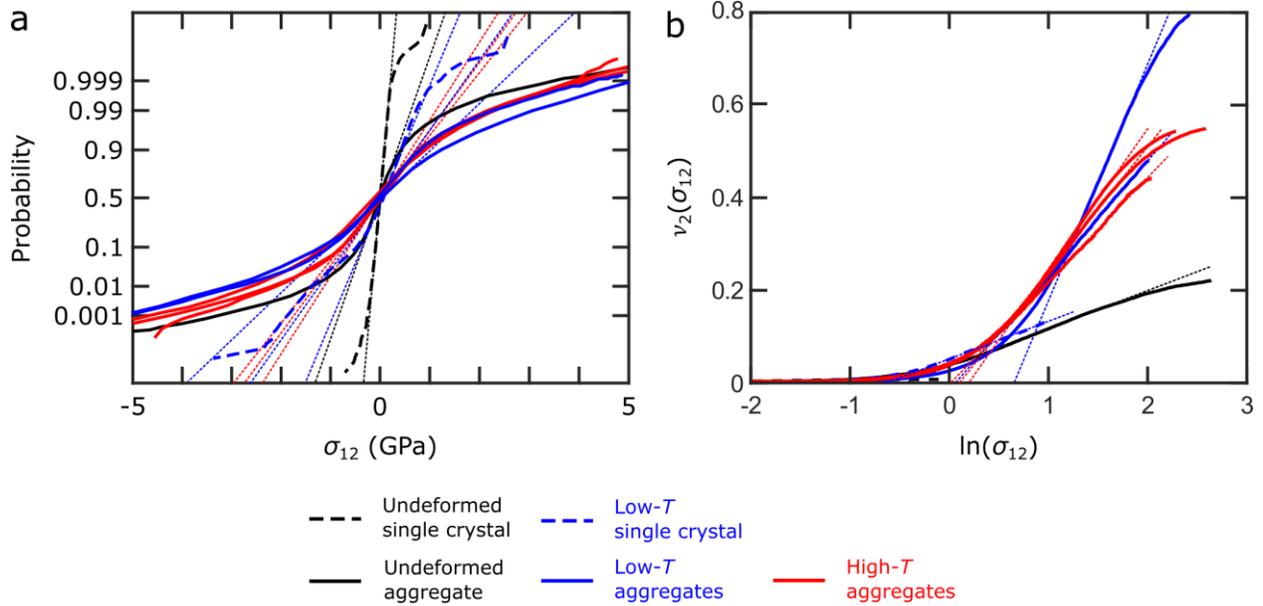

*Figure 4.* (a) Normal probability plot of normalised $\sigma_{12}$. On this plot, normal distributions fall on a straight line (i.e., fine dotted lines), therefore, deviations from straight lines indicate departures from normal distributions. Solid and dashed lines are calculated from HR-EBSD data. (b) Restricted second moments of the normalised $\sigma_{12}$ probability distributions. On this plot, straight lines (i.e., fine dotted lines) indicate that probability $P(\sigma_{12}) \propto \sigma_{12}^{-3}$, as expected of stress fields generated by dislocations[43–46]. Solid and dashed lines are calculated from HR-EBSD data.

The plot of restricted second moments of the stress distributions in Figure 4b characterises the form of the tails of the probability distributions (Methods). All samples exhibit significant portions of their probability distribution that fall on straight lines in Figure 4b, indicating that $P(\sigma_{12}) \propto \sigma_{12}^{-3}$, as expected of stress fields generated by dislocations[43–46] (Methods). Several of the curves depart from straight lines at the highest stresses due to averaging of the elastic strains over the finite volume illuminated by the electron beam[44]. In this plot, the curves corresponding to the deformed single crystal and aggregates exhibit steeper gradients than those of their corresponding control samples, consistent with the presence of greater dislocation densities in the deformed samples (Methods). The aggregates deformed at 25°C exhibit distributions comparable to those deformed at 1150–1250°C in both Figures 4a and 4b. The greatest stress heterogeneity is present in sample San372b, one of the aggregates deformed at 25°C.

## *Autocorrelation of residual-stress fields*

Figure 5 presents the autocorrelation functions of normalised $\sigma_{12}$ in each sample (Methods). Sample MN1 exhibits a relatively flat autocorrelation function with negligible long-range correlation in its stress field. All other samples exhibit peaks in their autocorrelation functions with correlation lengths on the order of 1 μm. Sample San382t, a single crystal, exhibits pronounced anisotropy in its autocorrelation function, corresponding to the banded structure in its stress field. All of the aggregates exhibit isotropic autocorrelation functions.

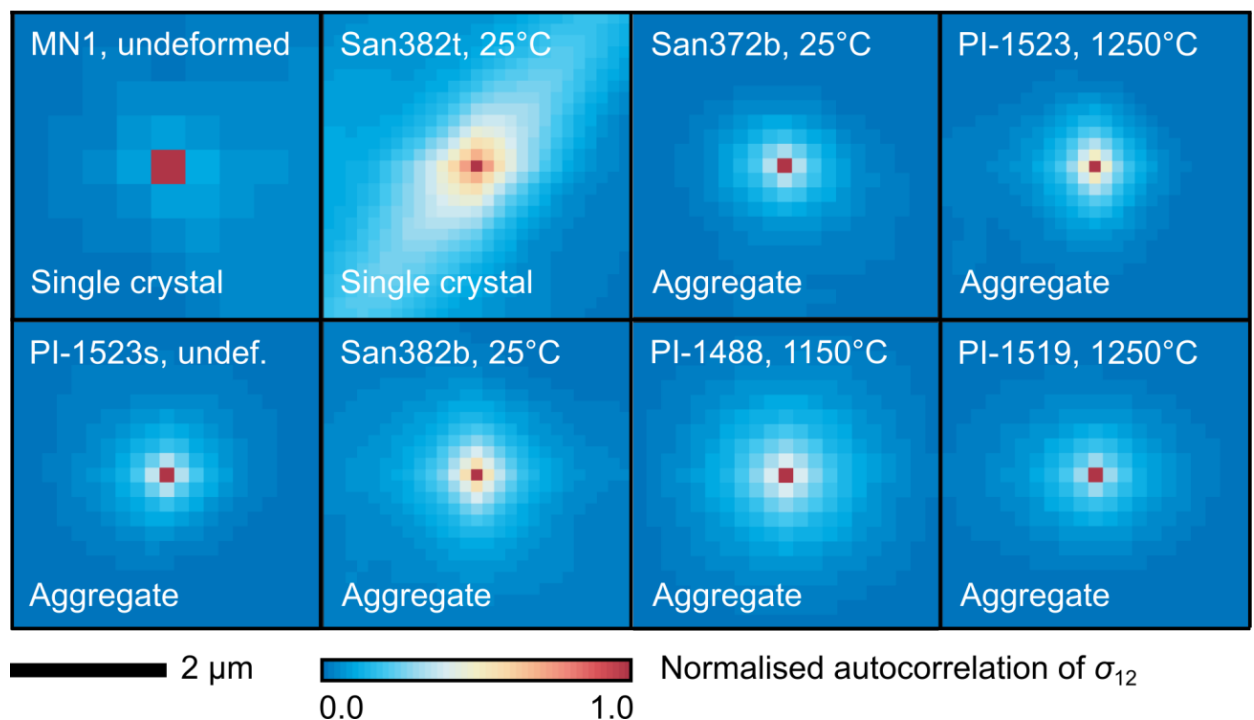

*Figure 5. Autocorrelation functions of normalised $\sigma_{12}$. Autocorrelation functions are normalised to unit peak height. The central $2 \times 2$ µm of the autocorrelation function is plotted for each sample. Annotations indicate the sample number, final deformation temperature, and whether the sample is a single crystal or an aggregate.*

## Discussion

The results reveal several new first-order characteristics of the residual stress fields of deformed olivine. Whilst Wallis et al.[36] previously reported stress heterogeneities on the order of 1 GPa in samples deformed at room temperature (including San382t and San382b), single crystals of olivine deformed at 1000°C and 1200°C exhibited stress heterogeneities with lesser magnitudes, on the order of a few hundred megapascals[37]. A surprising result of the present study is that, in contrast to single crystals, the aggregates of olivine deformed at 1150–1250°C exhibit stress heterogeneity with magnitudes again frequently on the order of 1 GPa, closely comparable to the aggregates deformed at room temperature (Figures 2–4).

One hypothesis for the cause of increased stresses in aggregates deformed at high temperature is that they are imparted by thermal contraction (whereas stresses imparted by the 300 MPa of decompression should be small relative to stresses observed in the aggregates decompressed by 6.9–9.3 GPa). However, the sample isostatically hot pressed at 1200°C and cooled without high-temperature deformation exhibits less stress heterogeneity than the samples deformed at high temperatures (Figures 2–4). Moreover, detailed analysis of the probability distributions of stress heterogeneity indicate that the high-magnitude stresses are imparted by dislocations. The stress distributions of samples deformed at both low and high temperatures exhibit high-stress tails that deviate from normal distributions (Figure 4a) and are typical of materials deformed by dislocation-mediated mechanisms, even at low temperatures[36,43–45]. Importantly, the analysis of restricted second moments of the probability distributions indicates that the tails of the distributions follow $P(\sigma_{12}) \propto \sigma_{12}^{-3}$ in all the deformed samples, as expected of stress fields generated by dislocations[43–46] (Methods). The similarities among the probability distributions of residual stress in samples deformed at low and high temperatures, and the particular form of those distributions, provide strong evidence that the high stress magnitudes recorded by the samples deformed at high temperatures are also predominantly imparted by the dislocation content.

This interpretation is consistent with the previous analysis of single crystals of olivine deformed at high temperatures[37]. In those samples, the stress heterogeneity was clearly controlled by the dislocation content rather than constraints imposed by neighbouring grains (as is also the case for San382t[36]). The single crystals were deformed at similar final macroscopic differential stresses (218 MPa and 388 MPa) to the aggregates deformed at high temperatures in this study (204–258 MPa, Table 1), and therefore the *total* dislocation densities of the two sets of samples should be similar[48]. However, in contrast, the GND densities observed here (commonly > $10^{14}$ m$^{-2}$, Figure 1) within the aggregates deformed at high temperatures are orders of magnitude higher than those in the single crystals deformed at high temperatures (on the order of $10^{12}$ m$^{-2}$)[37]. The difference in GND densities is consistent with well-developed theory for the grain-size dependence of GND density[49–51], in which strain-compatibility constraints imposed by neighbouring grains cause a greater fraction of the total dislocation density to manifest as GND density in finer grained materials. As GNDs impart significant non-cancelling stress fields, differences in GND density explain the observed differences in the magnitudes of stress heterogeneity between single crystals and aggregates deformed at both low and high temperatures (Figures 2–4)[37].

The interpretation that the residual stress distribution is modified by GND density is consistent with the role of GNDs in generating long-range internal stresses[51,52]. Correlation lengths on the order of 1 µm (Figure 5), along with the spatial distributions evident in the stress maps (Figure 2), demonstrate the presence of long-range internal stresses in all the deformed samples. Similar characteristic length scales of stress heterogeneity occur in the single crystals of olivine deformed at high temperature[37] and around nanoindents placed in single crystals of olivine at room temperature[36,39]. Therefore, long-range internal stresses are increasingly apparent as a ubiquitous characteristic of olivine deformed by dislocation-mediated mechanisms. Future work should address the scaling of long-range internal stresses at lower strain rates and even higher temperatures, under which conditions subgrain-boundary formation and (sub)grain boundary migration are more effective at reducing GND density in subgrain interiors.

The presence of long-range internal stress has been explicitly linked to the transient mechanical behaviour of samples deformed at low temperatures. Both single crystals and aggregates of olivine deformed at room temperature exhibit a Bauschinger effect and, in particular, the relative magnitudes of strain hardening and yield stresses during cyclic loading demonstrate that hardening is dominantly kinematic[31]. These interpretations of the mechanical data are consistent with the presence of long-range internal stresses preserved in the samples (Figures 2–5)[36].

Long-range internal stresses also play a key role in establishing steady-state creep by dislocation-accommodated grain boundary sliding, which was inferred by Hansen et al.[23] to be the dominant deformation mechanism in the aggregates deformed at high temperatures that we have investigated here. During dislocation-accommodated grain boundary sliding, strain compatibility among grains is maintained by dislocation motion in grain interiors, with the rate of dislocation motion limiting the macroscopic strain rate[23,53]. The observed mechanical behaviour (in terms of sensitivity to applied stress and grain size)[23] is consistent with the predictions of microphysical models in which glide is impeded by generation of back stresses among dislocations as they pile up against subgrain boundaries[53]. Our new observation of long-range internal stresses generated by the dislocation content in these samples is consistent with this model (Figures 2–5). The mounting evidence that back stresses among dislocations control strain hardening at low temperatures and steady-state creep at high temperatures raises the question of whether they also make a significant contribution to transient creep at high temperatures.

Current models for transient creep are commonly formulated based on stress transfer between weak and strong slip systems to maintain strain compatibility in an aggregate of mechanically anisotropic grains[12,29]. The initial increments of strain, upon application of macroscopic differential stress, are likely to be accommodated by grains well oriented for glide on the weakest slip system. Due to compatibility constraints, continued deformation on the weakest slip system progressively transfers load to the stronger slip systems until they are sufficiently active to result in steady-state creep. This model was originally formulated for ice[12] and has since been applied to olivine[29], and has inspired recent formulations that are incorporated into large-scale models of postseismic deformation[1,2]. Those formulations describe the viscosity evolution based on combinations of the steady-state flow laws for single crystals of different orientations[1,2]. However, it is important to recognise that, along with aggregates, single crystals deformed at high temperatures also exhibit strain-hardening transient creep (except when limited by the rate of dislocation multiplication, which results in strain softening)[9,33,34]. Therefore, transient creep of an aggregate includes some contribution from intragranular processes that is not captured in the steady-state flow laws typically incorporated in descriptions of postseismic transient creep. At the small strains involved in postseismic deformation (often on the order of microstrain), the transients caused by intragranular processes (i.e., hardening of each slip system) are likely important throughout the postseismic interval but certainly dominate the very earliest deformation that precedes the transfer of stresses among grains. However, previous attempts to model transient creep of single crystals of olivine have been largely limited to fitting phenomenological models due to a lack of constraints on the underlying microphysical processes[9].

Our results provide a major step forward by highlighting the relevant microphysics associated with transient creep of olivine aggregates. We suggest that the storage and release of back stresses among dislocations provides a conceptual basis for the development of a new generation of models for the contribution of intragranular processes to transient creep of olivine aggregates at high temperatures. This hypothesis is based on the role of back stresses among dislocations in generating strain hardening at low temperatures[31,36], the role of back stresses in steady-state deformation at high temperatures[23,53], and the observation of stress heterogeneity with similar magnitudes and length scales generated by the dislocation content in both sets of samples (Figures 2–5). Importantly, although postseismic stress changes affect rocks in both low- and high-temperature regimes in the lithosphere and asthenosphere, our analyses indicate that the same microphysics controls transient behaviour in both cases. This commonality offers potential for a model of transient deformation that is applicable across a variety of settings and conditions. Hansen et al.[31] calibrated a set of constitutive equations for transient low-temperature plasticity based on storage and release of back stress among dislocations. This formulation therefore provides a basis for modelling transient creep at high temperatures if the impact of additional recovery mechanisms, such as dislocation climb, can be taken into account.

# Conclusions

Aggregates of olivine deformed at temperatures of 1150–1250°C exhibit intragranular stress heterogeneity that is remarkably similar in both magnitude (Figures 2–4) and correlation length (Figure 5) to aggregates deformed at room temperature. The form of the probability distributions of internal stress (Figures 3 and 4), along with comparable distributions (albeit of lesser magnitude) in single crystals (Figures 2–4)[37], indicate that the high-magnitude stresses are generated primarily by the dislocation content. The difference in magnitudes of stress heterogeneity between single crystals and aggregates is consistent with the role of GNDs, which occur in higher densities in the aggregates (Figure 1), in generating long-range internal stress[50–52]. These observations contribute to a growing body of evidence[31,36,37] suggesting that the storage and release of back stresses among dislocations may be a significant cause of transient creep that is currently not incorporated in models of postseismic deformation. The formulation of new models for these intragranular processes will refine the predictions of large-scale geodynamic simulations.

# Methods

## *Deformation experiments*

We utilise samples deformed in two sets of experiments described in detail by Hansen et al.[23,31]. Key information on the deformation conditions is summarised in Table 1. As the deformation conditions typically varied during the experiments (e.g., to obtain flow-law parameters), we report the final conditions in Table 1.

Samples San382t and San382b were deformed at room temperature whilst paired together in a single assembly in a deformation-DIA apparatus[31]. High confining pressures inhibited fracture, whilst differential stresses of several gigapascals induced deformation by low-temperature plasticity. This sample was subjected to shortening followed by extension and exhibited strain hardening of several gigapascals in both cases[31]. The microstructures, including stress heterogeneity, lattice misorientation, and distributions of dislocations, of these samples have been characterised in detail by Wallis et al.[36]. The samples contain straight dislocations, often arranged in slip bands, that are spatially associated with stress heterogeneity. Sample San372 was deformed in a similar manner but subjected only to shortening, rather than cyclic loading[31].

Samples PI-1488, PI-1523, and PI-1519 were deformed at final temperatures of 1150°C and 1250°C and confining pressures of 300 MPa in a gas-medium Paterson apparatus[23]. Based on the mechanical data and microstructures, Hansen et al.[23] inferred that these samples deformed by dislocation-accommodated grain boundary sliding. Whilst this deformation mechanism involves sliding on grain boundaries, the majority of strain is accommodated by dislocation motion in grain interiors. The samples contain abundant subgrain boundaries and more diffuse lattice misorientation[23].

We also analyse two samples that act as undeformed controls. The first is a single crystal of San Carlos olivine. The mapped area lacks subgrain boundaries and the crystal contains a low density, $< 10^{10}$ m$^{-2}$, of free dislocations[47]. The second is a portion of the starting material, sample PI-1523s, used for experiment

PI-1523. The specimen was extracted after sample fabrication by isostatic hot pressing of powdered San Carlos olivine at 1200°C but prior to the deformation experiment.

**Table 1.** Sample details

| Sample | Grain size[1] (μm) | Final temperature (°C) | Final confining pressure (GPa) | Final differential stress (GPa) | Final strain rate (s$^{-1}$) | Finite plastic strain (%) | Notes |
|---|---|---|---|---|---|---|---|
| MN1 | N/A | N/A | N/A | N/A | N/A | 0 | Undeformed single crystal |
| PI-1523s | 6.5 | 1200 | 0.3 | N/A | N/A | 0 | Aggregate isostatically hot pressed in Paterson apparatus |
| San382t | ~700 | 25 | 6.9 | -3.8 | $2.7 \times 10^{-5}$ | 7.2 | Single crystal deformed by cyclic loading in a D-DIA apparatus |
| San382b | 3.0 | 25 | 6.9 | -3.8 | $2.7 \times 10^{-5}$ | 5.3 | Aggregate deformed by cyclic loading in a D-DIA apparatus |
| San372b | 4.6 | 25 | 9.3 | 4.3 | $1.2 \times 10^{-5}$ | 2.1 | Aggregate deformed by unidirectional loading in a D-DIA apparatus |
| PI-1488 | 9.6 | 1150 | 0.3 | 0.258 | $0.9 \times 10^{-5}$ | 5 | Aggregate deformed in Paterson apparatus |
| PI-1523 | 5.4 | 1250 | 0.3 | 0.257 | $8.9 \times 10^{-5}$ | 17 | Aggregate deformed in Paterson apparatus |
| PI-1519 | 4.6 | 1250 | 0.3 | 0.204 | $9.7 \times 10^{-5}$ | 19 | Aggregate deformed in Paterson apparatus |

[1]Grain sizes were measured using the line intercept method on EBSD data with a stereological scaling factor of 1.5[23,31].

## *Sample preparation and data acquisition*

Deformed samples were sectioned parallel to the loading axis and prepared for electron backscatter diffraction (EBSD) analysis. The cut surfaces were ground flat and polished with successively finer diamond down to a grit size of 0.25 μm. The surfaces were finished with either 0.05 μm diamond or 0.03 μm colloidal silica. Samples San382t and San382b were coated with 0.5 nm Pt/Pd. The remaining samples were left uncoated.

EBSD data were acquired in two field emission gun scanning electron microscopes (SEMs) equipped with Oxford Instruments AZtec acquisition software and NordlysNano detectors. Samples San382t and San382b were analysed under high vacuum in a Philips XL-30 SEM at Utrecht University. The remaining samples were analysed at low vacuum (50 Pa of water vapour) in an FEI Quanta 650 SEM at the University of Oxford. Table 2 presents the details of each dataset. Reference frame conventions were validated following the approach of Britton et al.[54] and the microscopes were calibrated for high-angular resolution electron backscatter diffraction (HR-EBSD) following the approach of Wilkinson et al.[40].

**Table 2.** Dataset details

| Sample | Number of map points | Step size (μm) | Number of pixels in diffraction patterns |
|---|---|---|---|
| MN1 | 400 × 500 | 0.5 | 1344 × 1024 |
| PI-1523s | 450 × 650 | 0.2 | 1344 × 1024 |
| San382t | 155 × 295 | 0.15 | 1344 × 1024 |
| San382b | 320 × 200 | 0.15 | 1344 × 1024 |
| San372b | 300 × 450 | 0.2 | 640 × 480 |
| PI-1488 | 575 × 850 | 0.2 | 640 × 480 |
| PI-1523 | 116 × 172 | 0.2 | 1344 × 1024 |
| PI-1519 | 600 × 800 | 0.2 | 640 × 480 |

## *High-angular resolution electron backscatter diffraction*

We mapped densities of geometrically necessary dislocations (GNDs) and intragranular stress heterogeneity using HR-EBSD postprocessing of the diffraction patterns. This technique measures lattice rotation and elastic strain heterogeneity by using cross correlation to track small shifts in features within EBSD patterns[39–41]. After the cross-correlation procedure, we filtered out results for which the normalised peak in the cross correlation function was < 0.3 and those with a mean angular error in the deformation gradient tensor > 0.004[42]. GND densities were estimated from the spatial gradients in the lattice rotations using the approach of Wallis et al.[36]. Elastic strain and hence stress are measured relative to the strain state at a reference point chosen within each grain. In deformed materials, it is likely that no part of the material is free from elastic strain, so the results provide maps of relative intragranular strain/stress heterogeneity. To provide an intuitive metric that is independent of the choice of reference pattern, we normalise the stresses by subtracting the mean value of each component within each grain[37,55,56]. Therefore, the resulting maps document stress heterogeneity relative to the mean stress state of each grain, which we refer to simply as normalised stresses, **σ**, herein. We focus on the $\sigma_{12}$ component as this component is least modified by sectioning the samples and generally most closely pertains to the glide forces on dislocations during deformation.

*Probability analysis of stress heterogeneity*

To characterise the probability distributions of the normalised stresses we use normal probability plots and calculate the restricted second moments of the probability distributions. On a normal probability plot, the probability axis is scaled such that a normal distribution falls on a straight line. Departures from a straight line indicate departures from a normal distribution and are commonly observed at high stress magnitudes in deformed materials. Work on Cu, InAlN, and steel has demonstrated that the stress fields of dislocations typically cause these high-stress 'tails' in the probability distributions[45,56]. Analysis of the restricted second moment of the probability distribution of normalised $\sigma_{12}$ offers a powerful means to test the causal relationship between the dislocation content and high-stress tails in the probability distributions. The restricted second moment, $v_2$, is a metric that characterises the shape of a probability distribution, $P(\sigma)$, based on the integral over restricted ranges in stress, calculated as

$$v_2(\sigma) = \int_{-\sigma}^{+\sigma} P(\sigma)\,\sigma^2\,d\sigma \tag{1}$$

,
[45]. A remarkable property of a population of straight parallel dislocations is that, regardless of their spatial configuration, the probability distribution of its stress field tends to $P(\sigma) \propto \sigma^{-3}$ at high stresses, following

$$P(\sigma) \to C\rho\sigma^{-3} \tag{2}$$

,
where $C$ is a constant that depends on the material, type of dislocation, and considered stress component, and $\rho$ is the dislocation density[43–46]. As a consequence, a plot of $v_2$ versus normalised $\sigma_{12}$ becomes a straight line at high stresses if the stress field exhibits the $P(\sigma) \propto \sigma^{-3}$ form expected of a population of dislocations, and the gradient of that line is proportional to the dislocation density[44,45]. Whilst $C$ can be determined for simple populations of dislocations (e.g., one type of edge dislocation)[43–45], no theory has been developed for more complex populations of dislocations. Therefore, we do not attempt to quantify $\rho$ in our samples, which contain multiple types of dislocations on different slip systems[23,36], oriented differently with respect to the sample reference frame to which the stress is referred. However, we do use the relative magnitudes of these gradients to imply an approximate ranking in the dislocation densities between samples.

To characterise the length scales of stress heterogeneity, we employ autocorrelation functions of normalised $\sigma_{12}$. Spatial correlation (e.g., stresses of the same sign likely being adjacent) generates a peak in the autocorrelation function. The width of the peak indicates the characteristic correlation length of the stress field.

## Acknowledgements

This work was funded by Natural Environment Research Council grant NE/M000966/1 and startup funding to D. Wallis from Utrecht University.

## Author contributions

D. Wallis and L.N. Hansen conceived the study and performed deformation experiments. D. Wallis acquired and processed the microstructural data and drafted the manuscript. All authors contributed to data interpretation and revision of the manuscript.